\documentclass[11pt]{article}
\usepackage[numbers,sort&compress]{natbib}	
\usepackage{hypernat}				
\usepackage{graphicx}		
\usepackage{amssymb}
\usepackage{simplemargins}
\setallmargins{1in}
\usepackage[american]{babel}
\usepackage{amsmath}
\usepackage{amsfonts}	
\usepackage{amsmath}
\usepackage{url}
\usepackage{multicol}
\usepackage[table]{xcolor} \definecolor{lightyellow}{HTML}{FFFFBB}
\usepackage[pdftex]{hyperref}
\hypersetup{
  unicode=false,
  pdffitwindow=true,
  pdfpagelayout=TwoPageLeft,
  pdfhighlight=/P,
  pdfmenubar=false,
  pdftoolbar=false,
  pdfstartview=FitV,
  pdfnewwindow=true,
  colorlinks=true,
  linkcolor=blue,
  anchorcolor=black,
  citecolor=blue,
  filecolor=black,
  menucolor=black,
  urlcolor=blue,
  breaklinks=true,
  pdftitle={The Social Force Model and its Relation to the Kladek Formula},
  pdfkeywords={Pedestrian, Crowd, Mass, Simulation, Dynamics, Viswalk, Social Force Model, Kladek}
  pdfsubject={Social Force Model, Pedestrian, Crowd, Mass, Simulation, Dynamics, Viswalk, Vissim, Kladek},
  pdfauthor={Kretz},
  pdfcreator={Tobias Kretz},
  pdfproducer={T. Kretz},
}

\fontfamily{cmss}\selectfont

\title{The Social Force Model and its Relation to the Kladek Formula}

\fontfamily{cmss}\selectfont

\author{Tobias Kretz, Jochen Lohmiller, Johannes Schlaich\\
PTV Group, Haid-und-Neu-Stra{\ss}e 15, D-76131 Karlsruhe, Germany\\
\texttt{\{First.Family\}@ptvgroup.com}\\
}%

\begin{document}
\fontfamily{cmss}\selectfont
\maketitle
\fontfamily{cmss}\selectfont
\abstract{
It was recently found that the Social Force Model of pedestrian dynamics in a macroscopic limit for 1d movement does not reproduce the empirically found inflection point of the speed-density relation. It could be shown that, however, a simple and intuitively comprehensible extension of the Social Force Model shows the inflection point. Motivated by this observation in this contribution the relation of the Social Force Model with the Kladek formula for the speed-density relation of urban motorized traffic is discussed. Furthermore the models are compared to results data from experiments on vehicular, cycling, and pedestrian dynamics.}

\fontfamily{cmss}\selectfont

\section{Introduction Part I: Empirical Data on Pedestrians' Speed-Density Relation}

\begin{figure}[ht!]
\centering
\includegraphics[height=5cm]{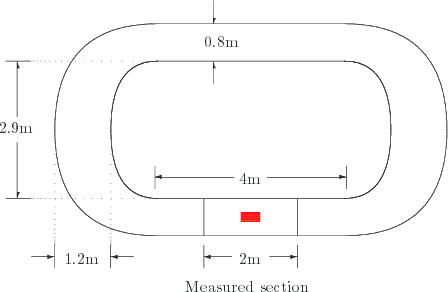} \hspace{12pt}
\includegraphics[height=5cm]{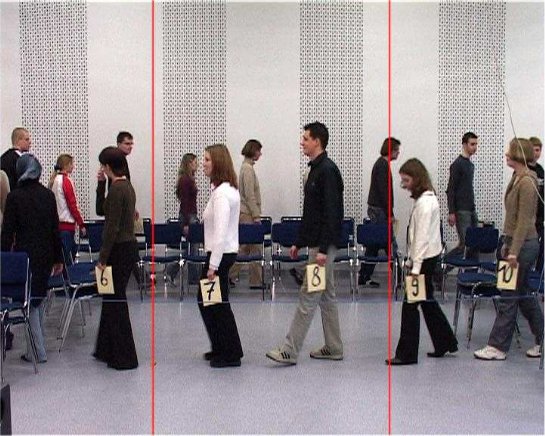}
\caption{Experimental setup. Source: Figures 2 and 3 of \cite{seyfried2005fundamental}}
\label{fig:ExperimentalSetup}       
\end{figure}

In the course of recent years a number of experiments have been conducted in which pedestrians walk single-file in a closed loop \cite{seyfried2005fundamental,chattaraj2009comparison,seyfried2010phase,seyfried2010enhanced,Portz2011analyzing}. Having a different number of pedestrians in the loop different densities are prepared. In a section of the loop density -- in this case: line density [persons / m] -- and speed are measured. Figure \ref{fig:ExperimentalSetup} shows the experimental setup from which most data stems. There were experiment runs in which the loop was larger and more pedestrians participated, but the principle was always the same.

\begin{figure}[ht!]
\centering
\includegraphics[height=6cm]{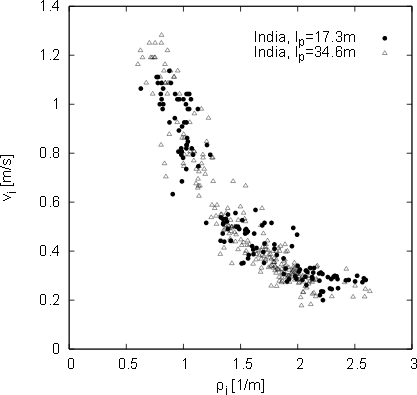}\hspace{12pt}
\includegraphics[height=6cm]{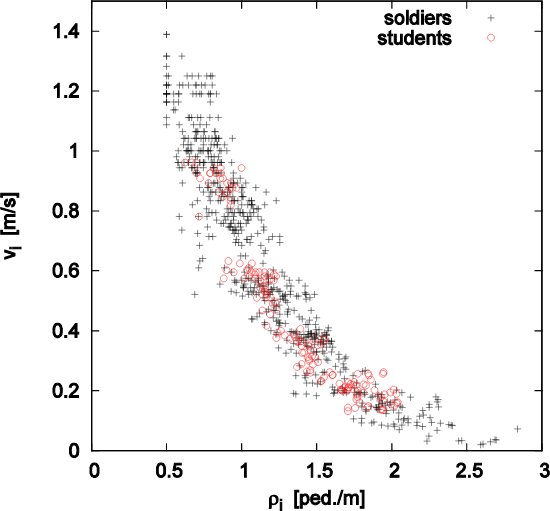}
\caption{Speed density diagram for different loop sizes and professions in India (left) and Germany (right). In India the participants in both experiments were students, in Germany the participants in the short loop experiment were students, the participants in the long loop were soldiers. Sources: left: Figure 7 of \cite{chattaraj2009comparison}; right: Figure 1 of \cite{Portz2011analyzing}}
\label{fig:shortlong}       
\end{figure}

The experiment has been conducted at various places around the world. Figure \ref{fig:shortlong} shows for example results for India and Germany for various loop sizes and different social background (professions) of participants. In all cases the curvature of an imagined function approximating the data has a negative curvature at small densities and -- more important and interesting -- a positive curvature at high densities. With the positive curvature at high densities speed approaches zero for increasing densities only ``slowly''. This is an important property of the speed density relations that have been found in these experiments. Models of pedestrian dynamics should be able to reproduce the change of the sign of the curvature with density, i.e. that the speed density relation has an inflection point. A simpler and necessary (although not sufficient) condition is that the curvature of the speed-density relation -- the second derivative of speed by density -- vanishes at some certain density.

\section{Introduction Part II: the Kladek Formula}
In 1966 Kladek \cite{kladek1966geschwindigkeitscharakteristik} suggested the following equation as macroscopic function to describe the relation between average momentary speed $\bar{v}_m$ and density $\rho$ of motorized urban road traffic:
\begin{equation}
\bar{v}_m = \bar{v}_f \left(1-e^{-\gamma\left(\frac{1}{\rho}-\frac{1}{\rho_{max}}\right)}\right) \label{eq:Kladek}
\end{equation}

``Average momentary speed'' -- also called ``space-averaged speed'' -- means that for speed-density data to comply the Kladek formula, speed has to be measured first, along an extended section of the road at one time and second by measuring how far a vehicle moves in an externally given, very short time span. The latter is sometimes forgotten, although it is equally important: if speed is measured by measuring the time it takes to move a fixed distance, speed zero cannot occur.

If speed is expressed in fraction $f$ of the free speed $\bar{v}_f$ and density as fraction $x$ of the maximum (i.e. stand still) density the Kladek formula reduces to
\begin{equation}
f(x) = 1-e^{-a\left(\frac{1}{x}-1\right)} \label{eq:Kladek-ggu}
\end{equation}
leaving only one free parameter -- $a$ -- to adjust the function to empirical data. The relation between parameter $a$ in equation (\ref{eq:Kladek-ggu}) and parameter $\gamma$ in equation (\ref{eq:Kladek}) is $\gamma = a \rho_{max}$.

In their text book Lohse and Schnabel \cite{lohse2011grundlagen} point out that the Kladek formula has a number of favorable properties in the sense that one single formula covers the entire density range giving free speed for zero density and zero speed for maximum density.

We note that the Kladek formula has an inflection point which is located -- for equation (\ref{eq:Kladek-ggu}) -- at $x_i=a/2$. 

\begin{figure}[ht!]
\centering
\includegraphics[height=4.7cm]{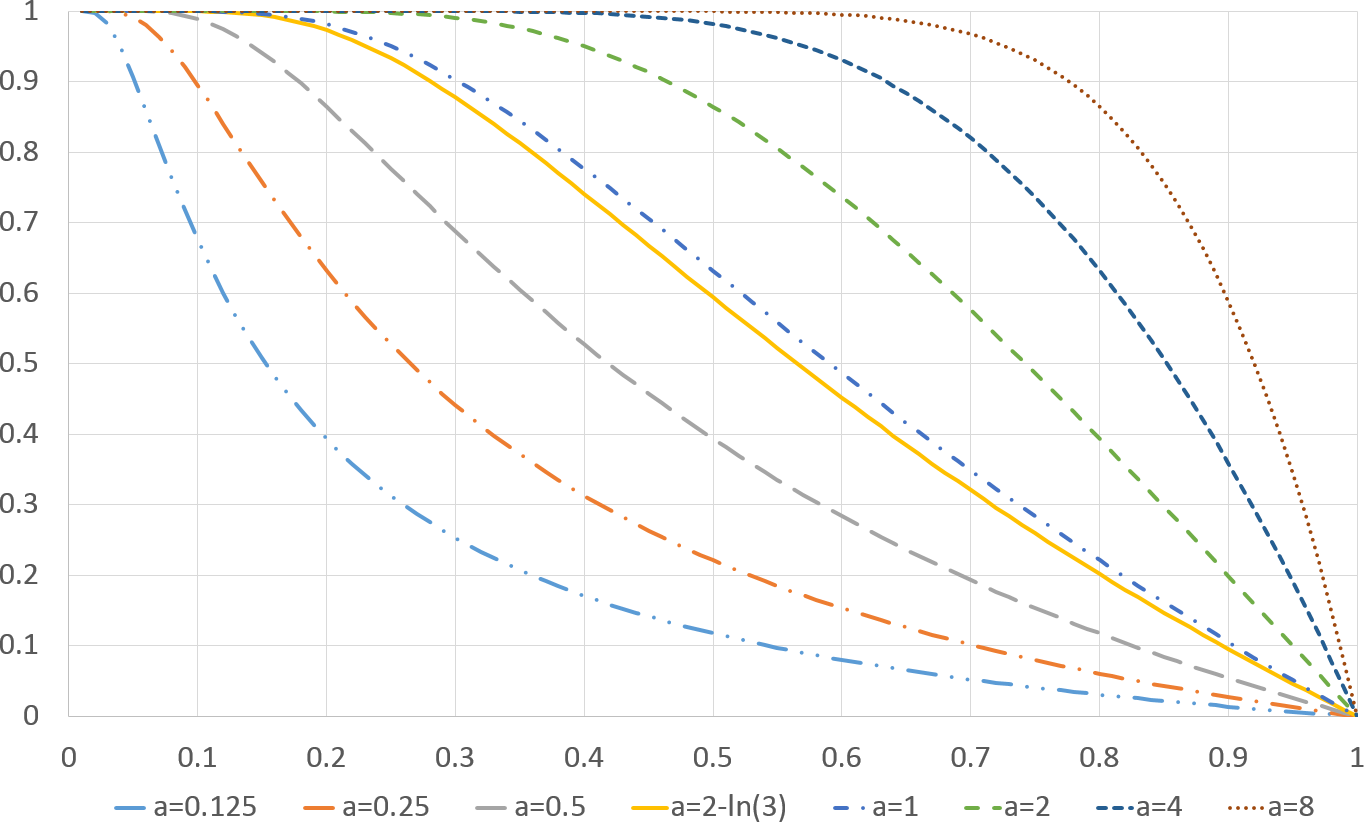} \hspace{10pt}
\includegraphics[height=4.7cm]{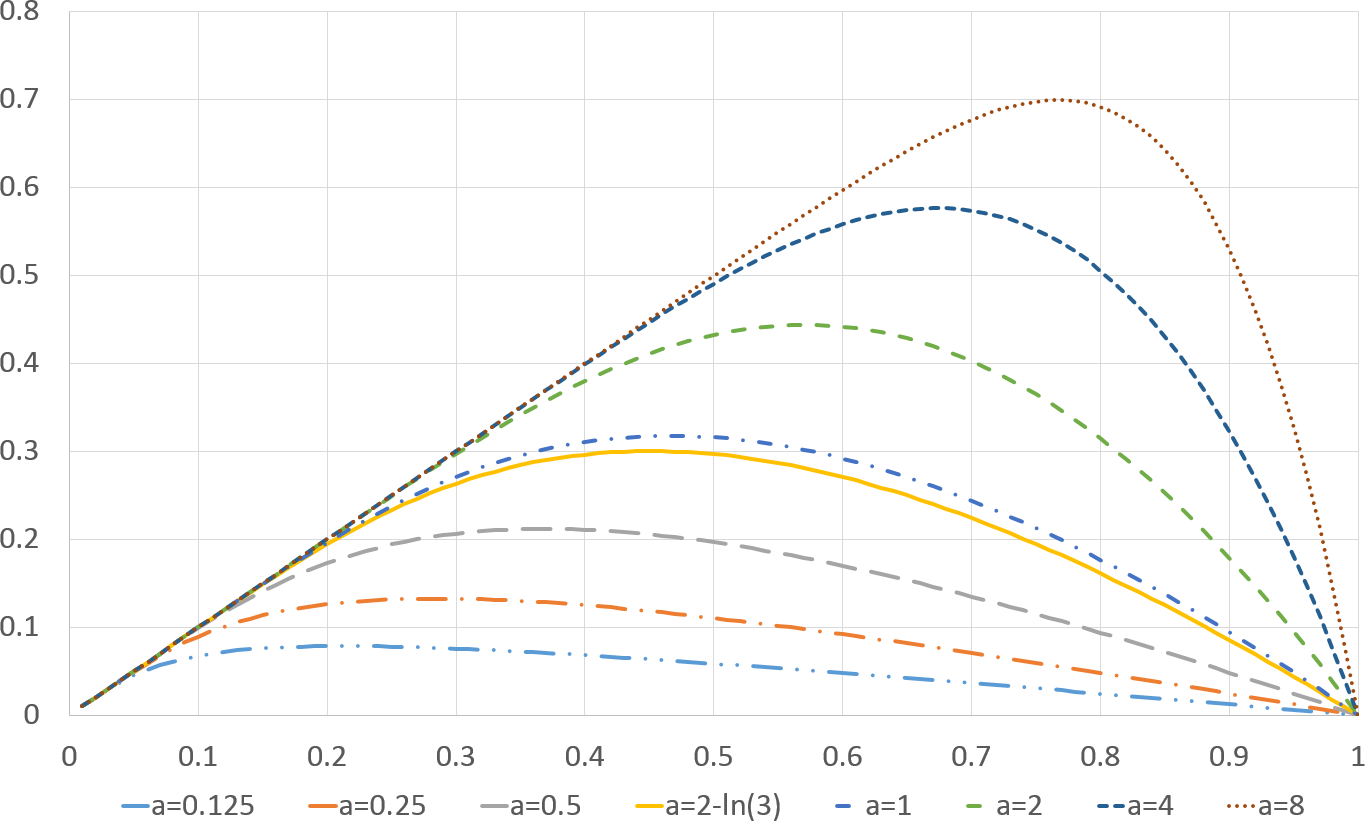}
\caption{Speed-density (left) and flow-density (right) relations according to equation (\ref{eq:Kladek-ggu}) for various values of parameter $a$.}
\label{fig:Kladek}       
\end{figure}

It is sometimes pointed out for road traffic that the maximum of the flow is located at or near the speed drop in the speed density relation \cite{wu2000verkehr,kuhne2004fundamentaldiagramm}. For vehicular traffic this has been achieved with piece-wise definitions of the fundamental diagram. For a smooth and analytical function as the Kladek formula this means that the inflection point of the speed-density relation (where the first derivative is most negative) has to be at the position of the maximum of flow\footnote{This is necessarily fulfilled, if the Gaussian function is used to describe the relation between speed and density, compare \cite{lohse2011grundlagen}.}. For equation (\ref{eq:Kladek-ggu}) this requirement translates into two equations:
\begin{eqnarray}
\frac{\partial}{\partial x} (x f(x))|_{x_c} &=& 0 \\
\frac{\partial^2}{\partial x^2} f(x)|_{x_c} &=& 0
\end{eqnarray}
which when solved fix the value of parameter $a$ to $a=2-\ln(3)\approx 0.901$, leaving no more freedom for calibration with empirical data. However, Figure \ref{fig:Kladek} which shows the speed density and the flow density relation for various values of parameter $a$, suggests that this value does not yield the closest approximation to empirical data.

Figure \ref{fig:Kladek} shows that for different values of parameter $a$ the maximum in the flow-density diagram can be on the left or on the right side of $x=0.5$. Since usually in pedestrian as well as vehicular traffic the maximum is located left of half the maximum density it is interesting to know the value of $a$ for which the maximum is exactly at $x=0.5$ and that is where 
\begin{equation}
(1+2a)=e^a
\end{equation}
therefore
\begin{equation}
a = -W_{-1}\left(-\frac{1}{2}e^{-\frac{1}{2}}\right)-\frac{1}{2} \approx 1.256
\end{equation}
with $W_{-1}()$ denoting the lower real-valued branch of the Lambert W function.

Weidmann applied the Kladek formula to describe the speed-density relation of uni-directional pedestrian flow \cite{weidmann1993transporttechnik,buchmueller2006parameters}. This led to the situation that -- despite Kladek introduced the equation to describe urban vehicular flow -- today the Kladek formula is and has been applied predominantly for pedestrian dynamics and engineering \cite{daamen2004modelling,bellomo2008modelling,lammel2008bottlenecks,lammel2009matsim,venuti2009crowd,beltaief2011multi,galiza2011estimating,papadrakakis2011footbridge,bruno2011non,bruckner2012combined,huth2012fundamental,lussmann2012network,zhang2012pedestrian,nikolic2014probabilistic,martinez2014reinforcement,chen2014towards} with only few literature search hits related to vehicular traffic \cite{daniel2009real}.

Weidmann's parametrization of the Kladek formula to describe uni-directional, two-dimensional pedestrian dynamics is
\begin{eqnarray}
v_f &=& 1.34 \text{ m/s}\\
\gamma &=& 1.913 \text{ 1/sqm}\\
\rho_{max} &=& 5.4 \text{ 1/sqm}
\end{eqnarray}

For the dimensionless form -- and therefore valid in a one- as well as two-dimensional setting -- the value of parameter $a$ would be
\begin{equation}
a = \frac{\gamma}{\rho_{max}} \approx 0.354
\end{equation}

\section{The Social Force Model for Steady-States in Single-File Movement}
It has recently been shown \cite{Kretz2016inflection} that in a macroscopic limit and for single-file movement the speed-density relation of the Social Force Model \cite{Helbing2000simulating,johansson2007specification} is

\begin{equation}
\dot{x}_\alpha = v_0 - (1-\lambda) \tau A \frac{1}{e^{\frac{1}{B\rho}}-1} \label{eq:steadystatedensity}
\end{equation}
where $\rho$ is the line density of pedestrians.

It was furthermore shown that if not each pedestrian exerts a force on every other, but only nearest neighbors effect mutually, then the resulting speed-density relation is
\begin{equation}
\dot{x}_\alpha = v_0 - (1-\lambda) \tau A e^{-\frac{1}{B\rho}} \label{eq:steadystatedensityNN}
\end{equation}

We note that equation (\ref{eq:steadystatedensityNN}) is different from equation (\ref{eq:Kladek}) only through parametrization. The functional form is identical. The parameter equivalents are
\begin{eqnarray}
v_0&=&v_f\\
\frac{1}{B}&=&\gamma\\
\frac{(1-\lambda)\tau A}{v_0}&=&e^{\frac{\gamma}{\rho_{max}}}
\end{eqnarray}

\pagebreak[3]

We note as a result of this work:\nopagebreak 
\begin{table}[h]\nopagebreak 
\centering\nopagebreak 
\label{tab:highlight2}\nopagebreak        
\rowcolors{1}{lightyellow}{lightyellow}\nopagebreak 
\begin{tabular}{|p{9cm}|}
\hline
If the circular specification or the elliptical specification II of the Social Force Model are modified such that only nearest neighbors exert a force onto each other, then the speed density relation for homogeneous steady-state one-dimensional movement is given by the Kladek formula. Therefore the Kladek formula is a macroscopic limit of a modified Social Force Model.
\\ \hline 
\end{tabular}
\end{table}

Since the Kladek formula has an inflection point this implies that also the nearest-neighbors Social Force Model in its macroscopic limit shows an inflection point. 

In \cite{Kretz2016inflection} an extension of the Social Force Model was proposed: the force from the second next pedestrian should be suppressed by a factor $k$ with $0\leq k < 1$, the force from the third next pedestrian by a factor $k^2$ and so on. This suppression would be independent of actual distance between pedestrians and would be determined only by  neighborhood rank. With this extension the resulting macroscopic speed-density relation of the Social Force Model is
\begin{equation}
\dot{x}_{k \alpha} = v_0 - (1-\lambda) \tau A \frac{1}{e^{\frac{1}{B\rho}}-k} \label{eq:steadystatedensityk}
\end{equation}
respectively in dimensionless form:
\begin{equation}
f(x)_k(\rho) = 1 - \frac{e^{a}-k}{e^{\frac{a}{x}}-k} \label{eq:steadystatedensityk-ggu}
\end{equation}

This shows that with parameter $k$ there is a continuous transformation from the original Social Force Model ($k=1$) to the Kladek formula ($k=0$), since
\begin{equation}
v_0 - (1-\lambda) \tau A e^{-\frac{1}{B\rho}} = v_0 - (1-\lambda) \tau A \frac{1}{e^{\frac{1}{B\rho}}-0}.
\end{equation}
where the left side is the Kladek formula in Social Force Model parametrization as given in equation (\ref{eq:steadystatedensityNN}).

\begin{figure}[ht!]
\centering
\includegraphics[height=4.4cm]{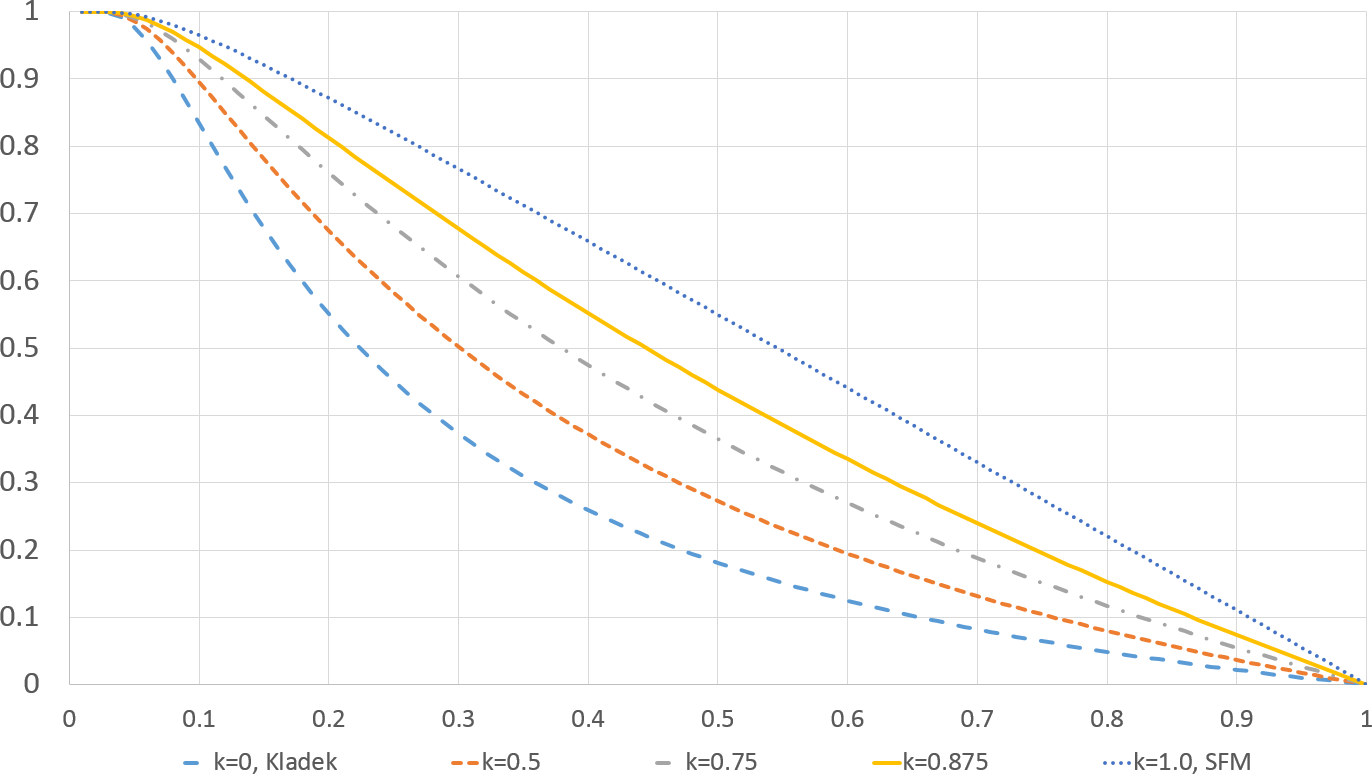} \hspace{10pt}
\includegraphics[height=4.4cm]{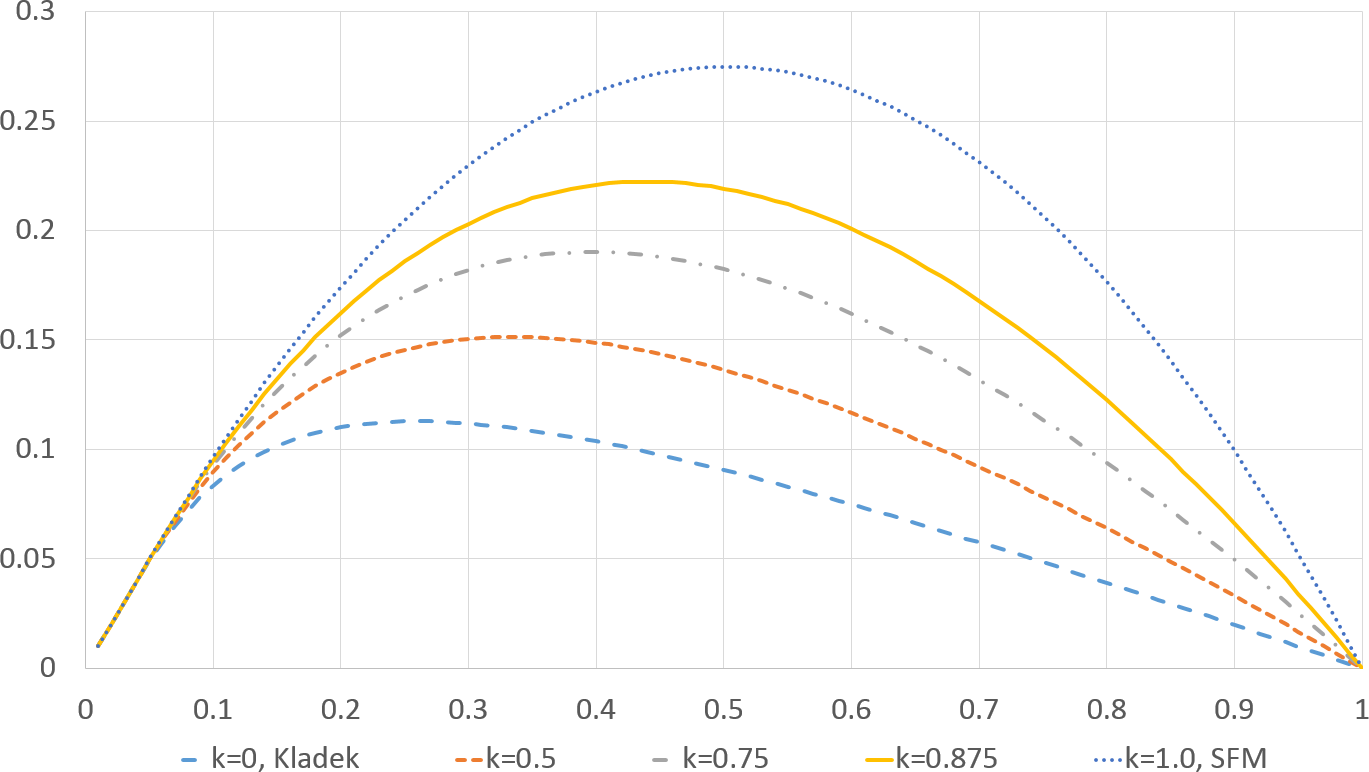}
\caption{Speed-density (left) and flow-density (right) relations of the $k$-extended Social Force Model for various values of parameter $k$. The plots are based on the dimensionless variants of equations and it is $a=0.2$.}
\label{fig:SFMk}       
\end{figure}

Fortunately for all values of $k<1$ there is an inflection point as can easily be seen in Figure \ref{fig:SFMk}. Thus all values $0<k<1$ can be seen to produce both, a modification of the Kladek formula as well as a modification of the Social Force Model, thereby preserving favorable properties like the existence of an inflection point.

Further favorable properties of this model extension are that it can be combined with arbitrary specifications of the Social Force Model or in general force-based models as for example \cite{yu2005centrifugal,yu2007modeling,chraibi2010generalized,campanella2014nomad} and that the new parameter $k$ re-opens the possibility for calibration even if one requires that the maximum of the flow be at same density as the inflection point in the speed-density relation.

\section{Comparison to Empirical Data}
\begin{figure}[hptb]
\centering
\includegraphics[width=0.75\textwidth]{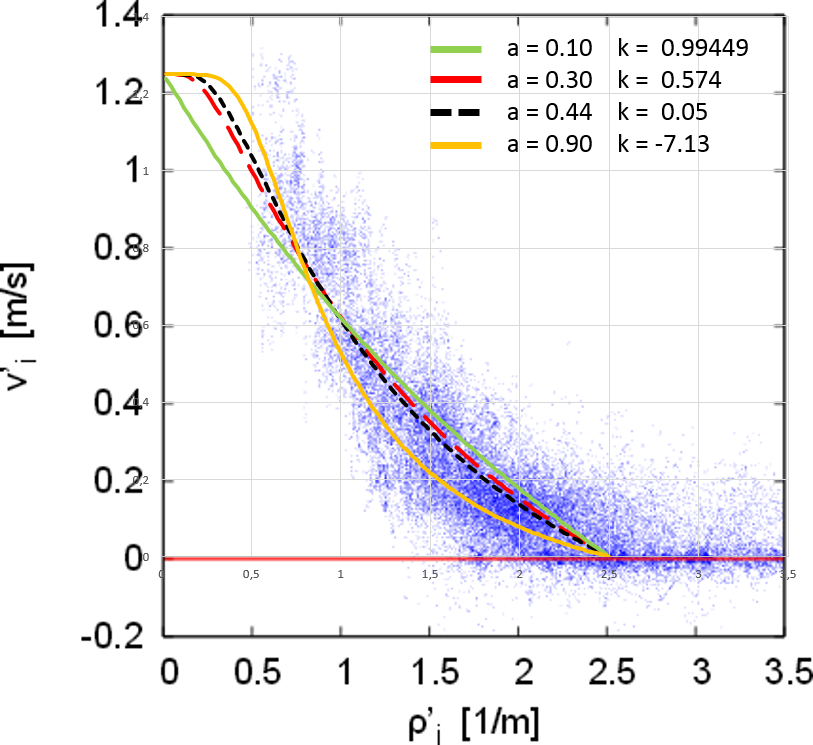}
\caption{Four functions with different parameters displayed on the background of Figure 3 of \cite{seyfried2010phase} which was produced with the same data as Figure \ref{fig:shortlong} but a different evaluation method. Free speed and maximum density were not calibrated, but were fixed a priori to $v_0=1.25$ m/s and $\rho_{max}=2.5$ $m^{-1}$.}
\label{fig:calibration-01}       
\end{figure}

\begin{figure}[hptb]
\centering
\includegraphics[width=\textwidth]{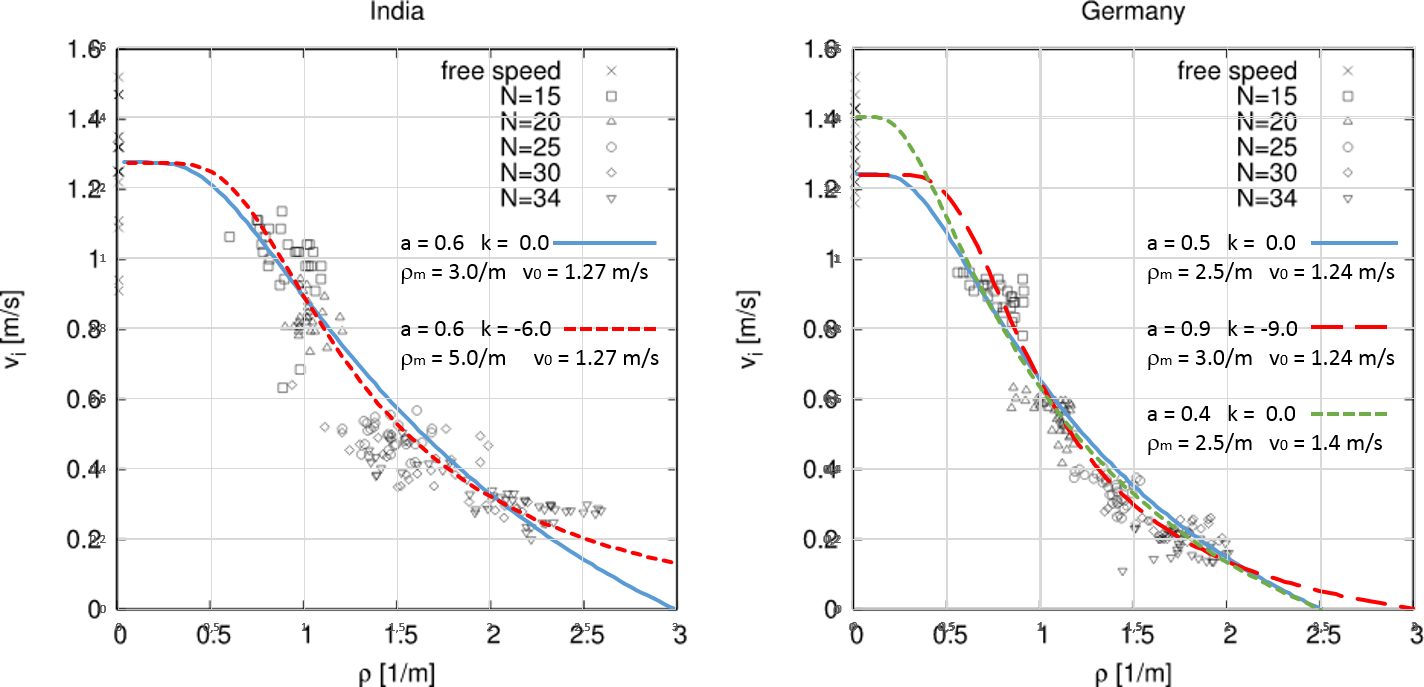}
\caption{Various functions overlaid with data recorded in India (left) and Germany (right). Source of original diagrams: Figure 5 of \cite{chattaraj2009comparison}. Especially for the data from India really small speeds do not occur -- possibly as a consequence of the measurement method -- and standstill density could be estimated only very roughly.}
\label{fig:calibration-02}       
\end{figure}

\begin{figure}[hptb]
\centering
\includegraphics[width=0.75\textwidth]{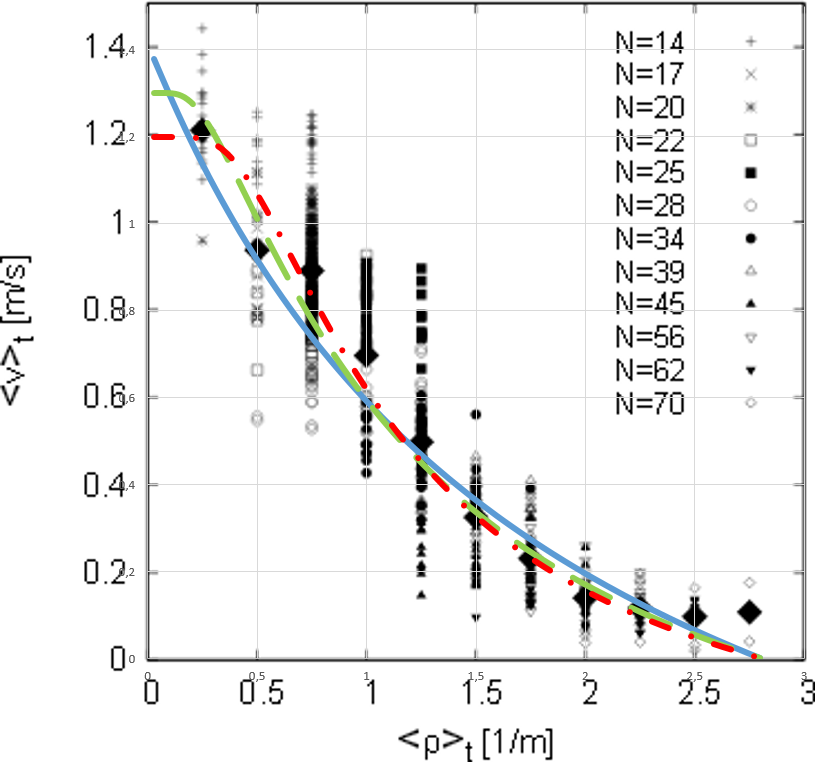}
\caption{Speed density diagram created from the same data as in Figure \ref{fig:shortlong} (right side, soldiers) overlaid with SFMk functions. For evaluation of the raw data a snap shot of all participants in the whole loop is taken at one point in time (resp. two frames) and speeds are measured whereas for density the global average density is used, see Figure 6 of \cite{seyfried2010enhanced}. The overlaid functions have the following parameters: solid blue line: $a=0.01$, $k=0.985$, $v_0=1.4$ m/s; dashed green line: $a=0.3$, $k=0.2$, $v_0=1.3$ m/s; dash-dotted red line: $a=0.5$, $k=-1.5$, $v_0=1.2$ m/s. Maximum density in all cases $\rho_{max}=2.8$ $m^{-1}$. Parameters of the blue and red curve have deliberately been pushed to extreme values to demonstrate the relation of values for $a$ and $k$ as well as the dependence on a choice for $v_0$.}
\label{fig:measurementvariants}       
\end{figure}

The theory so far was straight forward and one might expect that calibration of the resulting equations at empirical data should be equally straight forward. However, there are three issues: first, the theory assumes nonchalant that the average free speed and the maximum density are known. Yet measuring a meaningful average speed near maximum density is  difficult if not impossible since one is faced with stop-and-go waves in pedestrian \cite{Portz2011analyzing} as well as vehicular traffic. Second, the wide area covered by the speed-density pair data points makes a direct purely ``cloud-based'' -- just by eye-sight comparison -- calibration difficult. Numerous functions can appear equally well fitting and Figure \ref{fig:calibration-01} shows that these functions may have quite different parameters that determine their (similar) shape. And finally there is only few data available that has been collected and averaged in the momentary (space-averaged) way. The speed data in Figures  \ref{fig:shortlong} and \ref{fig:calibration-02} has been obtained by measuring the time it took pedestrians to walk a fixed distance. In this way very low speeds are systematically underrepresented compared to a measurement where speed is measured by measuring the distance pedestrians move in a fixed time and speed zero can never occur in contrast to Kladek's formula. Figure \ref{fig:measurementvariants} shows space-averaged data, but the publication does not state clearly if the speeds were measured with fixed $\Delta t$ or fixed $\Delta x$. Definitely a problem is that a measurement was triggered each time a pedestrian crossed position $x=0$ which means that if density was so high that no one could move no measurement was triggered. This implies that -- although data is space-averaged -- small speeds are systematically suppressed compared to when measurement is triggered externally. Since overtaking is not possible in these scenarios one can hope that the impact of the averaging method is nevertheless small. Still one has to bear the stated limitations in mind when interpreting the results of this section.

As mentioned there is a range of parameters that produce decent agreement of the macro functions with micro data. In Figure \ref{fig:calibration-01} one of the four functions even has a negative value for parameter $k$. Mathematically this is no problem, but it does not make sense with regard to the Social Force Model since then each second pedestrian would have an attractive and each other pedestrian a repelling effect.

Recently a comparison has been done between the fundamental diagrams of one-dimensional, uni-directional car, bike, and pedestrian traffic \cite{zhang2013comparative,seyfried2014universalities}. For this the already referenced data for pedestrians has been used, new data for cycling dynamics has been gathered and existing data from an experiment on vehicular dynamics \cite{sugiyama2008traffic,nakayama2009metastability} have been used. According to the description of measurement and (re-)evaluation here speed has been measured in exact accordance with the definition for momentary (quasi-instantaneous, space-averaged) measurement.

\begin{figure}[hptb]
\centering
\includegraphics[width=0.75\textwidth]{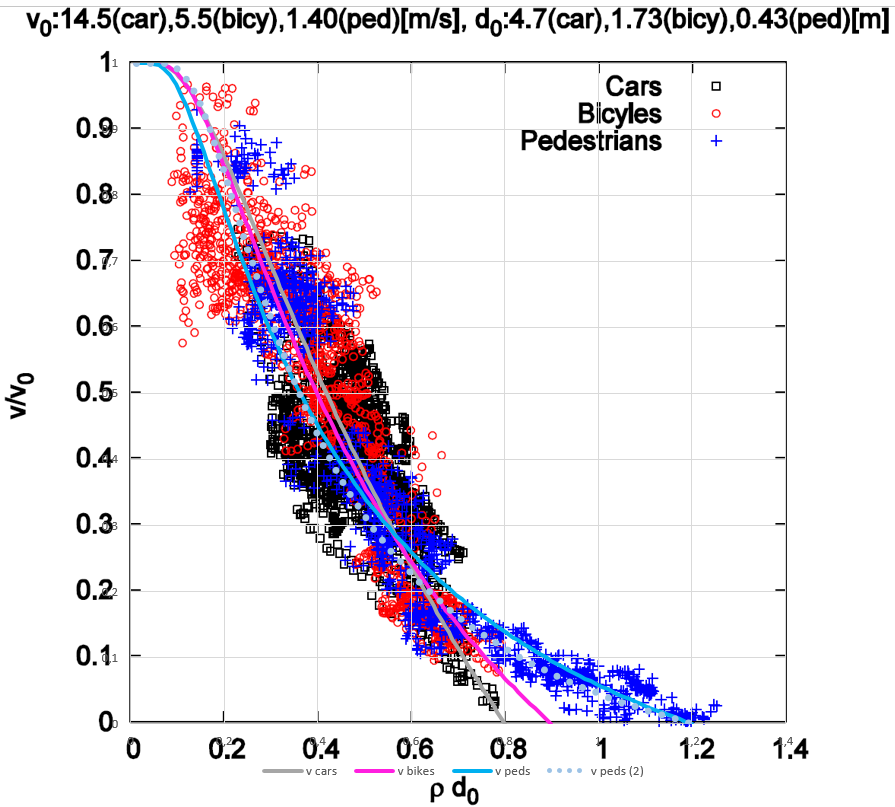}
\caption{Figure 7b from \cite{zhang2013comparative} overlaid with SFMk functions. ``density=1'' marks full occupancy (212.8 cars/km corresponding to $d_0=4.7$ m per car; 578 bikes/km, corresponding to $d_0=1.73$ m per bike; 2325.6 pedestrians per km, corresponding to $d_0=0.43$ m per pedestrian) and ``speed=1'' marks the maximum speed (cars: 14.5 m/s; bikes: 5.5 m/s and pedestrians: 1.4 m/s). Thus different to our approach in the paper density=1 marks the theoretically maximum density not the actual one. Please note the comments in \cite{zhang2013comparative} why densities larger 1 have been measured for pedestrians. The gray curve is our approximation to car data. It has $a=0.5$, $k=0.7$, and $x_{max}=0.8$. The magenta curve is our approximation to bike data. It has $a=0.5$, $k=0.2$, and $x_{max}=0.9$. The cyan curve is our approximation to pedestrian data with the restriction that it has to be $k\geq0$. It has $a=0.3$, $k=0.0$, and $x_{max}=1.2$. The dotted light blue is our approximation to pedestrian data without restriction for $k$. It has $a=0.5$, $k=-2.0$, and $x_{max}=1.2$. Maximum speed is 1 for all curves.}
\label{fig:calibration-carbikeped}       
\end{figure}

From Figure \ref{fig:calibration-carbikeped} it appears that $k_{car}>k_{bike}>k_{ped}$. One may conclude that car drivers look more ahead (react more to the next to next and further leading vehicle) than cyclists and those more than pedestrians do. This might be counter-intuitive since for car drivers the view to all but the immediate leader is occluded most. And this would not be without irony: the Kladek formula ($k=0$) fits better to describe pedestrian dynamics although it was introduced for vehicular (urban) traffic while the Social Force Model ($k=1$) describes rather vehicular dynamics although it was suggested as a model of pedestrian dynamics. This is or would be a nice a posteriori justification for those approaches where the Social Force Model was used to simulate vehicular traffic \cite{fellendorf2012social,schonauer2012modeling,rudloff2013comparing,anvari2014shared,anvari2014long,anvari2015modelling} and it is a motivation to work further in that direction.

However, given all the limitations of currently available data all this is preliminary and must be treated with care.

\section{Discussion -- Conclusions -- Outlook}
In this contribution we pointed out that with appropriate parameters an extended Kladek formula of the speed-density relation is the macro limit for an extended Social Force Model. Since before Weidmann used Kladek's formula to summarize approximate a lot of empirical data on pedestrian dynamics, this is an indication that both -- the Social Force Model as well as the Weidmann-Kladek speed-density relation -- bear some truth. At the same time it is a challenge for both that the Weidmann-Kladek formula and the Social Force Model are linked with two different values for parameter $k$.

The comparison with empirical data showed that on one hand it is possible to find parameters with which the extended Kladek formula fits well with the empirical data. On the other hand the parameters that result pose some questions. A reevaluation of existing trajectory data with different evaluation procedures might clarify this. If this would confirm that parameter $k$ requires to have negative values and if this cannot be explained satisfyingly from effects like body sway or technical artifacts of measurement and evaluation further modifications or extensions of the Social Force Model could be required.

A big topic which was not even touched in this work is the issue of a 2d extension of the ideas presented in this work. One cannot just assume that it yields realistic results if in 2d pedestrians are sorted by distance and then parameter $k$ is applied with an exponent according to distance rank. Instead one can think of a multitude of possible ways to extend the model for 2d: pedestrians could be ranked by neighborhood degree in a Voronoi diagram; or they could be ranked not by distance but force strength; ranking could be done in fan slices; one could even think of a multi-level sorting: first rank by time to collision and then rank those by distance where there will not be a collision judged by extrapolation of dynamics to the future, etc.

It would also be interesting to ``modify the modification''. Parameter $k$ was introduced because from the way the issue was addressed its introduction appeared to be the simplest way to extend the existing model. However, from a microscopic perspective it appears simpler to not introduce parameter $k$ but instead just cut the sum of forces at a certain maximum number of pedestrians or -- slightly more complex -- cut the summation at a maximum number of pedestrians and assign weights to the force of those pedestrians which are considered according to their neighborhood relation. In fact, this is what is done in vehicle traffic modeling, when standard one-leader car-following models are extended to so called multi-anticipative car-following models \cite{Bexelius1968extended,Lenz1999multi,Treiber2006delays,hoogendoorn2006empirical,Hoogendoorn2007multi,Farhi2012multi,Hu2014extended,costeseque2015multi}\footnote{Also the Vissim model can be said to be a multi-anticipative extension of the 1974 (and 1999) models by Wiedemann \cite{wiedemann1974simulation}, since the simulated drivers by default look 4 (previously 2) vehicles ahead and adapt their speed to the minimum computed value \cite{Vissim1993,Vissim2008}.}. The consequences of a cut-off of the summation of forces at a maximum number of pedestrians are particularly relevant since this is usually done in computer implementations of the Social Force Model for pragmatic reasons (computation speed).

Finally the consequences of different ways to measure density, speed, and flow \cite{Steffen2010b,nikolic2014pedestrian} (which then could not be expected to match the Kladek formula) could be interesting, particularly, if they bear consequences for the two-dimensional case.

\section{Acknowledgments}
For the preparation of this contribution we used David Pritchard's Transportation Research Board template for LaTeX\cite{Pritchard2009Latex}.
%

\bibliographystyle{utphys2011}
\bibliography{tgfbibs}


\end{document}